\documentclass{article}

\usepackage{geometry}
\usepackage{lineno,hyperref}
\modulolinenumbers[5]
\usepackage{siunitx}
\usepackage{amsmath,amssymb,amsfonts,amsthm,amsopn,dsfont, mathrsfs}
\usepackage{graphicx}

\usepackage[numbers]{natbib}
\usepackage{subcaption}
\usepackage{float}

\usepackage{bookmark}

\usepackage[tableposition=top]{caption}
\floatstyle{plaintop}

\usepackage{amsmath}
\usepackage{amssymb}
\usepackage{algorithm}
\usepackage{algorithmic}
\renewcommand{\vec}[1]{{\bf #1}}

\newcommand{\revOne}[1]{{\color{red} #1}}
\newcommand{\revTwo}[1]{{\color{blue} #1}}

\renewcommand{\revOne}[1]{#1}
\renewcommand{\revTwo}[1]{#1}

\begin{document}

\title{3D-1D modelling of \revOne{cranial mesh} heating induced by low or medium frequency magnetic fields}

\author{Alessandro Arduino\footnote{
  Istituto Nazionale di Ricerca Metrologica, 10135 Torino, Italy
  (a.arduino@inrim.it, o.bottauscio@inrim.it, u.zanovello@inrim.it, l.zilberti@inrim.it)} \and
Oriano Bottauscio$^*$ \and
Denise Grappein\footnote{
  Dipartimento di Scienze Matematiche, Politecnico di Torino, 10129 Torino, Italy (denise.grappein@polito.it, stefano.scialo@polito.it, fabio.vicini@polito.it)} $^,$\footnote{
  Istituto Nazionale di Alta Matematica F. Severi, 00185 Roma, Italy} \and
Stefano Scial\'o$^\dagger$ \and
Fabio Vicini$^\dagger$ \and
Umberto Zanovello$^*$ \and
Luca Zilberti$^*$
}

\date{August 13, 2025}

\maketitle
\begin{abstract}

  {\it Background and Objective:}
  Safety assessment of patients with one-dimensionally structured passive implants, like \revOne{cranial meshes} or stents, exposed to low or medium frequency magnetic fields, like those generated in magnetic resonance imaging or magnetic hyperthermia, can be challenging, because of the different length scales of the implant and the human body.
  Most of the methods used to estimate the heating induced near such implants neglect the presence of the metallic materials within the body, modeling the metal as thermal seeds.
  To overcome this limitation, a novel numerical approach that solves three-dimensional and one-dimensional coupled problems is proposed.

  {\it Methods:}
  The proposed method is compared with measurements performed on a \revOne{cranial mesh} exposed to the magnetic field generated by a gradient coil system for magnetic resonance imaging.
  Then, it is applied to a magnetic hyperthermia case study in which a patient with a \revOne{cranial mesh} is exposed to the magnetic field generated by a collar-type magnetic hyperthermia applicator for neck tumour treatment.

  {\it Results:}
  The experimental comparison of the proposed method predictions and the measurement data shows an improved accuracy \revTwo{near the maximum temperature increase} up to \SI{25}{\percent} with respect to the method based on thermal seeds.
  The application of the proposed method applied to the magnetic hyperthermia case study leads to a prediction of the maximum temperature increase that is \SI{10}{\percent} lower than the one overestimated by relying on thermal seeds.
  \revTwo{At the same time, the proposed method corrects the underestimation of the thermal seeds in the regions where the electromagnetic power is not directly deposited and the temperature increase is only due to heat transfer.}

  {\it Conclusions:}
  The proposed method leads to improved results with respect to previous approximations by modelling the thermal diffusion through the highly conductive metallic implants without affecting the anatomical human model discretization based on voxels.


\end{abstract}

\noindent\textit{The editorial version of the paper is available at doi:~\href{https://doi.org/10.1016/j.cmpb.2025.109009}{10.1016/j.cmpb.2025.109009}.}

\section{Introduction}
\label{sec:introduction}

\revOne{Medical technologies based on magnetic fields are, nowadays, common tools both in diagnostic and in therapeutic clinical practice.
A noteworthy example from diagnostics is magnetic resonance imaging (MRI)~\cite{brown2014}, which exposes the patient simultaneously to three different magnetic fields~\cite{gossuin2010}: a stationary field, typically equal to \SI{1.5}{\tesla} or \SI{3}{\tesla}; a radiofrequency (RF) field, whose frequency, typically of \SI{64}{\mega\hertz} or \SI{128}{\mega\hertz}, depends on the stationary field intensity; and a gradient field, whose time-dependent waveform has a frequency bandwidth ranging from about \SI{100}{Hz} up to a few kilohertz.}

\revOne{From the therapeutic side, a promising emerging technology is magnetic hyperthermia (MH) for cancer treatment~\cite{rajan2020}.
It is based on the experimentally observed positive effect of heat against cancer~\cite{datta2015}.
In MH, tissue heating is obtained through magnetic losses induced by an alternating magnetic field in magnetic nanoparticles (MNPs), which acts as thermal seeds, injected in the tumoural region.
The magnetic field applicators used in this context commonly operate in the low RF range, from \SI{100}{\kilo\hertz} to \SI{300}{\kilo\hertz}~\cite{rajan2020}.}

A critical step in the procedure that leads to the clinical adoption of medical technologies based on the exposure of patients to magnetic fields is their safety assessment.
For patients, the strict limitations introduced for workers and general public exposure in the ICNIRP guidelines~\cite{icnirp2010,icnirp2020} should not hold by default, because the benefit-risk principle in a clinical setting is different than in other scenarios. Nonetheless, regulatory agencies discipline the exposure of patients to fields generated by medical devices.
MRI equipment, for instance, is regulated by the IEC 60601-2-33 standard~\cite{iec2022}, which provides indications for all three fields to which the patients are exposed, to limit the possible occurrence of magnetophosphenes, temperature increase, nerve stimulation, and so on.
Regarding MH, instead, the definition of regulations is still ongoing and will deal with the injectable MNPs as well as the magnetic field applicator~\cite{rubia-rodriguez2021a}. For the latter, the Atkinson--Brezovich limit, forcing the product between the maximum intensity of the generated magnetic field $H$ and its frequency $f$ to comply with the constraint $H f < \SI{5e9}{\ampere\per\meter\per\second}$~\cite{hergt2007}, is currently adopted as a rule of thumb to guarantee no thermal stress in the tissues where the MNPs are not present. Nonetheless, evidences collected from the literature show that in certain cases this limit could be largely exceeded without inducing thermal damages outside the target region~\cite{angelakeris2017}.

Since the fourth edition of IEC 60601-2-33 has been published in 2022~\cite{iec2022}, some indications about patients with implantable medical devices have been included.
This addition became necessary because of the increasing number of medical implants within the population.
The presence of metallic, electrically conductive components in the patient's body could lead to complicated interactions with the magnetic field that should be taken into account to guarantee a safe MRI examination~\cite{winter2021}.
The complicated interaction could take place also with the magnetic field generated by the MH applicator, in which case the Atkinson--Brezovich limit does not provide a reliable safety condition.
To avoid unquantified risks, the presence of implants is currently used as an exclusion criterion for MH treatment, affecting a large portion of candidates~\cite{rubia-rodriguez2021a}.
In particular, in MH the risk originates from the Joule losses due to eddy currents induced by the magnetic field in the implant metallic components. The heat deposited in that way diffuses by thermal conduction from the implant to the surrounding biological tissues, causing an undesired temperature increase~\cite{rubia-rodriguez2021}.
The same physical effect is found in the interaction between a metallic implant and the gradient field of MRI~\cite{winter2021,arduino2021}, although currently it is regulated only for active implantable medical devices (AIMDs) by ISO/TS 10974~\cite{iso2018}. The feasibility of extending the prescriptions of this technical specification to passive implants is under investigation~\cite{arduino2022,bassen2022,zanovello2023}.

To assess safety conditions in presence of metallic implants, mathematical and numerical modelling are invaluable tools, which open the way to {\it in silico} trials~\cite{viceconti2021}.
In the context of electromagnetic dosimetry, {\it in silico} trials are particularly important, because they can predict power losses and thermal stresses generated within the human body, avoiding invasive and potentially dangerous measurements.
More specifically for medical devices, {\it in silico} trials were recently strongly legitimated by the first computational Medical Device Development Tool qualified by the US Food and Drug Administration (FDA)~\cite{fda2020}, to support MRI RF safety testing of AIMDs.

\revOne{Despite the large availability of powerful computational tools, also in commercial software, it remains a challenge to obtain reliable results in certain specific scenarios.
In particular, the topic of this paper is looking for a proper {\it in silico} assessment of thin (one-dimensionally structured) implants, which pose a challenge for their numerical modelling because of the different length scale of their small diameter (usually lower than \SI{1}{\milli\meter}) compared to the human body size.
It has been shown in the literature that good numerical solutions can be obtained by relying on non structured tetrahedral discretizations~\cite{golestanirad2017}.
However, heterogeneous anatomical human models are usually provided in voxel-based discretizations~\cite{kainz2019}, which can be directly obtained from MRI segmentations~\cite{iacono2013}.
A proper discretization of a thin metallic implant with voxels forces the adoption of extremely fine grids, leading to large computational costs and hardware resources~\cite{iacono2013}.}

\revOne{This paper will focus on the modelling of passive implants with a one-dimensional structure, like metallic stents~\cite{bottauscio2022} or metallic \revOne{cranial meshes}~\cite{bottauscio2023}.}
In this case, the presence of closed loops makes the implant subject to the induction of eddy currents when exposed to time-varying magnetic fields, like those generated by the MRI gradient coils (GCs) or the MH applicator.
Because of the large conductivity value of metal with respect to biological tissues, a proper estimation of the eddy currents induced by low or medium frequency magnetic fields can be obtained by taking into account only the implant and neglecting the surrounding tissues. This observation led to the recent proposal of a purely one-dimensional model based on a circuital description of the implant~\cite{bottauscio2023}.
Although this approach provides an accurate estimation of the Joule losses within the implant, a further step is needed to assess the thermal effects in the surrounding biological tissues.
To avoid the adoption of an extremely fine voxel-based discretization of the computational domain, recent publications dealing with this topic neglect the thin metallic implant in the thermal assessment~\cite{bottauscio2022,bottauscio2023}. Precisely, they model the implant as a set of thermal seeds, so that the previously computed Joule losses are distributed directly within the voxels of biological tissues. The temperature increase is finally computed according to Pennes' bioheat equation~\cite{pennes1948}.

Since the heating due to the Joule losses is not, in general, uniformly distributed within the metallic implant, the modelling approach based on thermal seeds is lacking of a physical contribution: thermal conduction through the implant itself. 
Being metals good thermal conductors, this contribution could significantly affect the final temperature distribution.
To take this contribution into account, relying on a voxel-based discretization of the biological tissues and keeping the computational burden reasonably low, numerical approaches addressing the coupling between three-dimensional and one-dimensional (3D-1D) problems on non-conforming discretizations can be adopted.
We here decide to adopt the optimisation-based domain decomposition approach recently presented in \cite{berrone2022a,berrone2022}.
Under this approach, auxiliary variables are introduced at the interface and a cost functional, mimicking the error committed in the fulfillment of interface conditions, is minimized constrained by the set of 3D-1D partial differential equations (PDEs). A similar optimisation-based 3D-1D coupling strategy was already applied successfully to the simulation of fluid and chemical exchanges in tumor-induced angiogenesis~\cite{berrone2023}.  Different strategies to address 3D-1D coupled problems can be found in the literature, such as in \cite{Dangelo2012}, relying on weighted Sobolev spaces, \cite{Zunino2019}, based on proper averaging operators, \cite{Koch2020}, resorting to kernel functions to approximate the singular behavior in a neighbourhood of the 1D inclusions, and \cite{Kuchta2021}, in which the 3D and the 1D problem are coupled by means of Lagrange multipliers. 

In this paper, the optimisation-based 3D-1D coupling strategy is used to assess the heating induced by exposing the metallic grid of a \revOne{cranial mesh} to a low or medium frequency magnetic field.
A comparison of the numerical results with experimental temperature measurements performed by exposing the \revOne{cranial mesh} to a system of MRI GCs is presented.
Finally, the proposed method is applied to a MH case study problem, in which a patient implanted with a \revOne{cranial mesh} undergoes a MH treatment for a neck superficial tumour.

\section{Methods}
\label{sec:methods}

\subsection{Mathematical and numerical modelling}
\label{sec:modelling}

\begin{figure}[t]
    \centering
    \includegraphics[width=9cm]{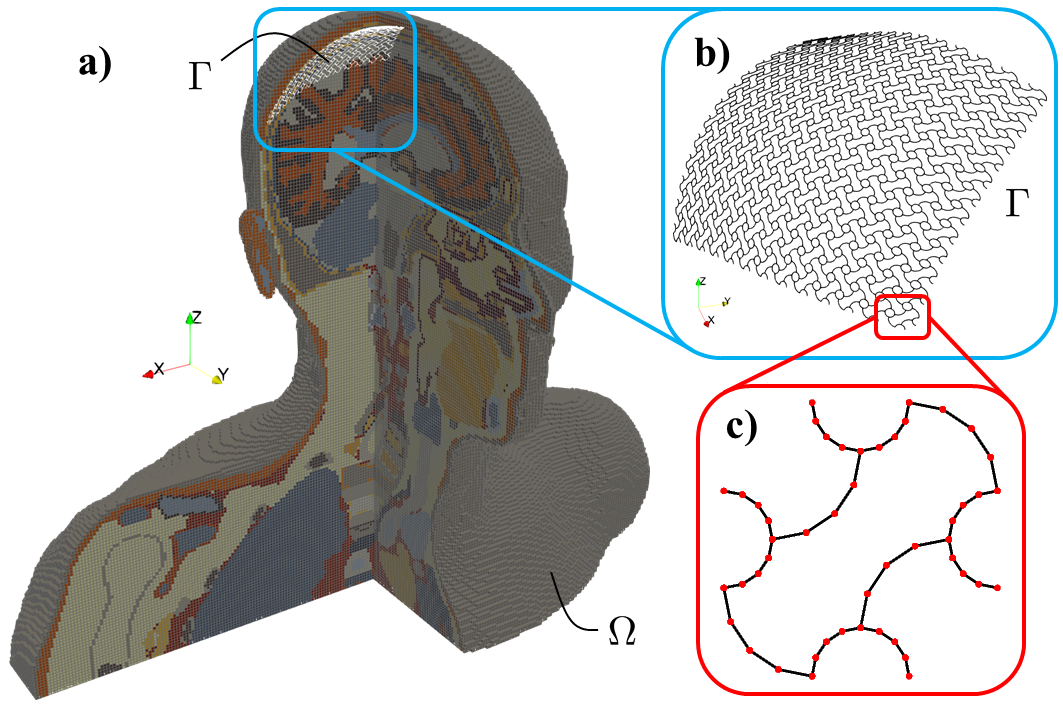}
    \caption{{\bf Sketch of a computational domain.} {\bf a)} The voxel-based anatomical human model Glenn, from the Virtual Population~\cite{gosselin2014}, is the computational domain $\Omega$. {\bf b)} The one-dimensional network $\Gamma$ models the \revOne{cranial mesh} implanted in Glenn. {\bf c)} The detail of the \revOne{cranial mesh} highlights the piecewise linear approximation in its geometry description (the red dots connect adjacent segments).}
    \label{fig:cranial-plate}
\end{figure}

The problem to be modelled is the heating of a filamentary medical implant with closed loops induced by a low or medium frequency magnetic field.
Hence, the computational domain is a portion of the human body, $\Omega\subset\mathbb{R}^3$, with an implanted medical device, $\Sigma\subset\Omega$. 
We assume that the filamentary implant $\Sigma$ is a network of thin tubular metallic wires with constant radius $r$.
Moreover, the implant is assumed to have a one-dimensional structure, in the sense that the radius $r$ is much smaller than the characteristic lengths of the implant itself and of the domain $\Omega$. 
As a consequence, each wire of $\Sigma$ can be approximated by its axis to get the one-dimensional network $\Gamma\subset\Sigma$, which will be used as the 1D representation of the filamentary implant in the following 3D-1D coupled model.
A sketch of the computational domain $\Omega$ and of the implant model $\Gamma$ is provided in Fig.~\ref{fig:cranial-plate} with reference to the case of a \revOne{cranial mesh}.

The problem involves two physical models solved in cascade. First, the electrical currents induced within the implant are evaluated through an electromagnetic model, previously presented in~\cite{bottauscio2023} and summarised in Section~\ref{sec:modelling:electromagnetic}. The resulting Joule losses are, then, used as forcing term of the thermal model, described in Section~\ref{sec:modelling:thermal}.

\subsubsection{Electromagnetic model}
\label{sec:modelling:electromagnetic}

It is assumed that the frequency of the magnetic field is sufficiently low so that: the secondary magnetic field due to the currents induced within the biological tissues is negligible with respect to the primary field; the induced displacement currents are negligible with respect to the induced conduction currents.
These are reasonable approximations for frequencies up to about \SI{1}{\mega\hertz}~\cite{bottauscio2023}.
Under this assumption, the large difference between the conductivity of metallic implants (of the order of \SI{1}{\mega\siemens\per\meter}) and the one of biological tissues (of the order of \SI{0.1}{\siemens\per\meter}~\cite{itis-database}) leads to eddy currents that are confined within the implant.
Therefore, the computational domain of the electromagnetic problem can be limited to just the one-dimensional model $\Gamma$ of the filamentary implant.

From the computational viewpoint, the assumed approximations allow one to handle $\Gamma$ as an electrical network, whose nodes (the red dots in Fig.~\ref{fig:cranial-plate}c) are connected by branches (the black segments in Fig.~\ref{fig:cranial-plate}c).
Each branch, representing a metallic wire, has a known resistance, deduced from its length, its cross section and the metal conductivity.
Moreover, each branch has, in series with the resistance, known self and mutual inductances with all other branches in the network. The inductive reactances can be non negligible, especially in the medium frequency range, and can be evaluated through numerical integration of the Biot-Savart law.
The induced electromotive forces in the network are represented by the vector potential of the source magnetic field, whose line integrals along the network branches provide the forcing term of the model.

To study the steady state behaviour of the network, usually reached after a transient response which is extremely short with respect to the characteristic time of the thermal phenomenon, phasor notation can be employed.
This leads to the description of the network model by means of a linear system whose unknowns are the induced electrical currents in the fundamental loops of the network, from which the current in each branch can be deduced.
Once the system is solved, assuming that $R$ and $I$ denote the resistance and the induced current, respectively, in a given branch of length $l$, the linear power density $p_{\rm em,l}$ dissipated in that branch can be estimated according to the Joule law as
\begin{equation}
\label{eq:joule-law}
p_{\rm em,l} = \frac{R I^2}{l}\,.
\end{equation}

The detailed mathematical derivation of the model is reported in a previous publication~\cite{bottauscio2023}.

\subsubsection{Thermal model}
\label{sec:modelling:thermal}

The heat generated in the medical implant due to the Joule losses diffuses by thermal conduction towards the biological tissues, whose thermal behavior is usually modelled according to Pennes' bioheat equation~\cite{pennes1948}. In temperature increase formulation, the bioheat problem reads~\cite{zilberti2014}
\begin{equation}
\label{eq:bioheat}
\left\{\begin{aligned}
&\rho c_{\rm p} \frac{\partial \vartheta}{\partial t} = \nabla \cdot \left( \lambda \nabla \vartheta \right) - h_{\rm b} \vartheta + p_{\rm em}\,,\quad{\rm in}\ \Omega\\
&\lambda \nabla \vartheta \cdot \vec{n} = - h_{\rm amb} \vartheta\,,\quad{\rm on}\ \partial\Omega
\end{aligned}\right.
\end{equation}
where $\vartheta$ denotes the temperature increase with respect to rest condition, $\rho$ is the mass density, $c_{\rm p}$ is the specific heat capacity, $\lambda$ is the thermal conductivity, $h_{\rm b}$ is the blood perfusion coefficient, $p_{\rm em}$ is the volume power density dissipated by the induced eddy currents, $h_{\rm amb}$ is the heat exchange coefficient with the surrounding environment, $\partial\Omega$ is the boundary of the computational domain $\Omega$ and $\vec{n}$ is its outward normal unit vector.
The value of each parameter is assumed to be constant.

The dissipated power density $p_{\rm em}$ originates from two contributions: the eddy currents induced within the biological tissues and the eddy currents induced within the implant.
The former are neglected in this paper, because the focus is on implant heating, but they could be added easily to the model for a complete safety assessment~\cite{bottauscio2023} when the magnetic field frequency is larger than \SI{100}{\kilo\hertz}~\cite{icnirp2020}.
The treatment of the eddy currents induced within the implant, instead, is a critical task and the main topic of this paper.
The direct simulation of a 3D model including both tissues and implant in the computational domain is, in general, unfeasible, as it would require a mesh resolution in the order of the micrometer~\cite{iacono2013}, leading to excessive computational costs and hardware requirements.
To avoid this issue, previous publications adopted a simple approach based on thermal seeds~\cite{bottauscio2022,bottauscio2023}.
A more accurate mathematical modelling, based on the 3D-1D coupling of the thermal phenomena in the biological tissues and in the metallic wires of the implant, is proposed in this paper.

\paragraph{Thermal seed approximation}
Each wire of the one-dimensional network $\Gamma$ modelling the filamentary implant is assumed as a thermal seed, namely the presence of the metal (with their thermal properties) is not modelled explicitly and the corresponding electromagnetic power is distributed within the tissues in which it is immersed~\cite{atkinson1984}.
According to the previously discussed electromagnetic model, each seed represents a line source with a uniform linear power density $p_{\rm em,l}$ provided by \eqref{eq:joule-law}.
From the practical viewpoint, a voxel-based discretization of the computational domain $\Omega$ is considered (like the one shown in Fig.~\ref{fig:cranial-plate}a) and the bioheat equation~\eqref{eq:bioheat} is solved numerically according to finite element method (FEM) with linear nodal shape functions.
In particular, the adopted implementation assumes that the forcing term $p_{\rm em}$ is constant within each voxel.
Hence, its value in a generic voxel is computed as the ratio between the total power dissipated in that voxel (obtained by summing up the power dissipated in the portion of each wire in $\Gamma$ crossing the voxel) and the voxel volume~\cite{bottauscio2023}.

Since, in general, the Joule losses are not uniformly distributed within the whole network $\Gamma$, the temperature distribution within the implant could be non-uniform and thermal conduction could take place through the implant itself, leading to a possible reduction of the peak temperature increase \revTwo{and the transfer of heat towards regions where the power is not directly deposited}.
This phenomenon cannot be modelled by the approach based on thermal seeds, which is therefore expected to overestimate the actual \revTwo{maximum} temperature increase \revTwo{and to underestimate the temperature increase elsewhere}.

\paragraph{3D-1D coupling}
To explicitly model the presence of the implant, the problem of interest can first be written as a 3D-3D coupled problem. We consider the temperature increase both in the implant $\Sigma$ and in the surrounding tissue $T$, where $\Omega=T\cup\Sigma$.
The boundary of $\Sigma$ and $T$ are denoted by $\partial \Sigma$ and $\partial T$, respectively, and $\partial T=\partial \Sigma \cup \partial \Omega$. The 3D-3D coupled thermal problem can hence be written as
\begin{equation}
\label{eq:bioheat3D3D}
\left\{\begin{aligned}
&\rho c_{\rm p} \frac{\partial \vartheta_{\tau}}{\partial t} = \nabla \cdot \left( \lambda \nabla \vartheta_{\tau} \right) - h_{\rm b} \vartheta_{\tau}\,,\quad{\rm in}\ T\\
& \rho_\sigma c_{{\rm p},\sigma} \frac{\partial \vartheta_{\sigma}}{\partial t} = \nabla \cdot \left( \lambda_\sigma \nabla \vartheta_{\sigma} \right) + p_{{\rm em},\sigma}\,,\quad{\rm in}\ \Sigma \\
&\vartheta_{\tau}=\vartheta_{\sigma}\,,\quad{\rm on}\ \partial\Sigma\\
&\lambda \nabla \vartheta_{\tau} \cdot \vec{n}_{\partial \Sigma}+\lambda_\sigma \nabla \vartheta_{\sigma} \cdot \vec{n}_{\partial \Sigma} = 0  \,,\quad{\rm on}\ \partial\Sigma\\
&\lambda \nabla \vartheta_{\tau} \cdot \vec{n}_{\partial \Omega} = - h_{\rm amb} \vartheta_{\tau}\,,\quad{\rm on}\ \partial\Omega
\end{aligned}\right.
\end{equation}
where quantities ${(\cdot)}_\sigma$ are defined inside $\Sigma$, and $\vartheta_{\tau}$ and  $\vartheta_{\sigma}$ denote the unknown temperature increases in $T$ and $\Sigma$, respectively. The equations written in the two domains are coupled by imposing the continuity of the temperature increase and the balance of heat fluxes at the interface $\partial \Sigma$.

The above formulation takes into account heat transfer phenomena within the implant. However, the discretization of problem \eqref{eq:bioheat3D3D} is computationally demanding, in particular for what concerns the definition of a computational mesh inside the thin wires of
$\Sigma$.
To overcome such a complexity, still retaining the advantage of the explicit representation of the implant, it is possible to operate a geometrical model reduction of the problem in $\Sigma$, yielding a 1D problem written on the network $\Gamma$. Along with the dimensional reduction of the domain $\Sigma$, the domain $T$ is replaced by the whole domain $\Omega$ and $p_{{\rm em},\sigma}$ is replaced by the previously computed $p_{\rm em, l}$, resulting in a 3D-1D coupled problem.

The mathematical formulation of such 3D-1D coupled problem in weak form is not standard. Here, only its discrete formulation is reported, based on FEM. The interested reader can refer to \cite{berrone2022a} for more details in a similar context.  \revOne{The key assumption that enables the model reduction is based on the fact that the total length $|\Gamma|$ of the filamentary implant, obtained by summing the length of all the wires, is much greater than its cross-sectional radius. For this reason, the error committed by assuming that $\vartheta_\sigma$ varies only along $\Gamma$ and not across the cross-sections of $\Sigma$ can be considered negligible.} 

Let $V_h$ and $\hat{V}_{\hat{h}}$ denote two standard finite element spaces, defined, respectively, on a voxel-based mesh in $\Omega$, with element diameter $h$, and on a 1D discretization of $\Gamma$, with element size $\hat{h}$. Let further $\revOne{Q^{\psi}_{\bar{h}}}$ \revOne{and $Q^{\phi}_{\bar{h}}$} be other finite element spaces on $\Gamma$, with  element size $\bar{h}$, possibly different from $\hat{V}_{\hat{h}}$. In a domain-decomposition framework, to decouple the problems in the tissue and in the implant, two auxiliary variables $\psi \in \revOne{Q^{\psi}_{\bar{h}}}$ \revOne{and $\phi \in Q^{\phi}_{\bar{h}}$} are introduced. In particular, variables $\phi$ and $\psi$ represent unknown functions at the interface between the two domains: $\phi$ is the unknown discrete flux and $\psi$ the unknown unique value of the discrete solution.
Then the discrete 3D-1D problem reads: \textit{find $(\vartheta_{h},\hat{\vartheta}_{\hat{h}}) \in V_h\times \hat{V}_{\hat{h}}$ and $\psi \in \revOne{Q^{\psi}_{\bar{h}}}$ \revOne{and $\phi \in Q^{\phi}_{\bar{h}}$}:}
\begin{align}
&\int_\Omega \rho c_{\rm p} \frac{\partial \vartheta_{h}}{\partial t} v_h \,{\rm{d}}V + \int_\Omega \lambda \nabla \vartheta_{h} \nabla v_h \, {\rm{d}}V  + \int_\Omega h_{\rm b} \vartheta_{h} v_h \, {\rm{d}}V +\nonumber \\
&\hspace{3cm} +
\int_{\partial\Omega} h_{\rm{amb}} \vartheta_h v_h {\rm{d}}S - \int_\Gamma 2\pi r\, \phi v_h \,{\rm{d}}L = 0 \,,\quad \forall v_h \in V_h \label{eq:3D1D_a}\\
&\int_\Gamma \pi r^2 \, \hat{\rho} \hat{c} \frac{\partial \hat{\vartheta}_{\hat{h}}} {\partial t}\hat{v}_{\hat{h}} \,{\rm{d}}L +\int_\Gamma \pi r^2 \,\hat{\lambda} \nabla \hat{\vartheta}_{\hat{h}} \nabla \hat{v}_{\hat{h}} \,{\rm{d}}L + \int_\Gamma 2\pi r \, \phi \hat{v}_{\hat{h}} \,{\rm{d}}L = \int_\Gamma \pi r^2 \,\hat{p}_{\rm em, l}\hat{v}_{\hat{h}} \,{\rm{d}}L \,,\quad \forall \hat{v}_{\hat{h}} \in \hat{V}_{\hat{h}} \label{eq:3D1D_b} \\
&\int_\Gamma \left(\vartheta_{h}-\psi\right)\eta \,{\rm{d}}L=0, \quad \forall \eta \in \revOne{Q^{\phi}_{\bar{h}}} \label{eq:3D1D_c} \\
&\int_\Gamma \left(\hat{\vartheta}_{\hat{h}}-\psi\right)\eta \,{\rm{d}}L=0, \quad \forall \eta \in \revOne{Q^{\phi}_{\bar{h}}} \label{eq:3D1D_d}
\end{align}
In the above equations, quantities $\hat{(\cdot)}$ are the restriction to $\Gamma$ of the corresponding quantities $(\cdot)_\sigma$ defined in $\Sigma$.
 The equations~\eqref{eq:3D1D_c}-\eqref{eq:3D1D_d} are the weak continuity condition, whereas flux balance is strongly enforced through the use of a unique flux variable $\phi$ for the 3D and 1D equations~\eqref{eq:3D1D_a}-\eqref{eq:3D1D_b}. 

Let us remark that, depending on the choice of the discretization spaces of the interface variables,  system~\eqref{eq:3D1D_a}-\eqref{eq:3D1D_d} could be badly conditioned, due to the non conformity of the mesh in $\Omega$ with respect to $\Gamma$.
A PDE constrained optimization method is proposed in \cite{berrone2022a} to alleviate ill conditioning issues. A functional is introduced to express the error in fulfilling the continuity condition, thus replacing equations~\eqref{eq:3D1D_c}-\eqref{eq:3D1D_d}. The solution is then obtained as the minimum of the functional constrained by equations~\eqref{eq:3D1D_a}-\eqref{eq:3D1D_b}.
The functional $J$ is defined as:
\begin{equation}\label{eq:functional}
    J(\phi,\psi)=\frac12\left(\|\vartheta_h(\phi)-\psi\|_\Gamma^2+\|\hat{\vartheta}_{\hat{h}}(\phi)-\psi\|_\Gamma^2\right),
\end{equation}
where $\|v\|_\Gamma$ is defined as:
\begin{displaymath}
    \|v\|_\Gamma= \left( \int_\Gamma v_{|\Gamma}^2 \,{\rm d} L \right)^{\frac{1}{2}}.
\end{displaymath}
The time-derivative is discretised using the backward Euler scheme, and the following optimization problem is solved at each time $t^k$, with time-step $\Delta t$:
\begin{align}
    &\min_{\phi^{k+1},\psi^{k+1}} J(\phi^{k+1},\psi^{k+1}) \ \  {\rm subject \ to:} \label{eq:3D1Dopt_t}\\
    &\int_\Omega \frac{\rho c_{\rm p}}{\Delta t}\left(\vartheta_h^{k+1}-\vartheta_h^{k}\right)v_h \,{\rm{d}}V + \int_\Omega \lambda \nabla \vartheta_h^{k+1} \nabla v_h \, {\rm{d}}V  + \int_\Omega h_{\rm b} \vartheta_h^{k+1} v_h \, {\rm{d}}V +\nonumber \\
&\hspace{1cm} +\int_{\partial\Omega} h_{\rm{amb}} \vartheta_h^{k+1} v_h {\rm{d}}S - \int_\Gamma 2\pi r\, \phi^{k+1} v_h \,{\rm{d}}L = 0 \,,\quad \forall v_h \in V_h \label{eq:3D1D_a_t}\\
&\int_\Gamma \frac{\pi r^2 \, \hat{\rho} \hat{c}}{\Delta t} \left(\hat{\vartheta}_{\hat{h}}^{k+1}-\hat{\vartheta}_{\hat{h}}^{k} \right)\hat{v}_{\hat{h}} \,{\rm{d}}L +\int_\Gamma \pi r^2 \,\hat{\lambda} \nabla \hat{\vartheta}^{k+1}_{\hat{h}} \nabla \hat{v}_{\hat{h}} \,{\rm{d}}L+ \nonumber \\
& \hspace{3cm} + \int_\Gamma 2\pi r \, \phi^{k+1} \hat{v}_{\hat{h}} \,{\rm{d}}L = \int_\Gamma \pi r^2 \,\hat{p}_{\rm em, l}\hat{v}_{\hat{h}} \,{\rm{d}}L \,,\quad \forall \hat{v}_{\hat{h}} \in \hat{V}_{\hat{h}} \label{eq:3D1D_b_t}
\end{align}

The algebraic form of the above problem can be obtained by simply collecting the integrals of basis functions into matrices.

\revOne{The numerical scheme used to solve the 3D-1D coupled problem is based on the approach described in \cite{berrone2022a} and \cite{berrone2022}. The variable $\hat \vartheta_\sigma$ was discretized using linear finite elements on a partition of $\Gamma$ made of 1D cells that are obtained as subdivisions of the segments discretizing the implant depicted in Fig.~\ref{fig:cranial-plate}c. A coarser mesh for the flux control variable $\phi$ was chosen to address issues related to system conditioning, as discussed in \cite{berrone2022a}. For the sake of simplicity, in this work, the same grid was used for the control variables $\phi$ and $\psi$. On this grid, $\psi$ was discretized with linear finite elements, while $\phi$ with piecewise constant basis functions.
The implementation of the method was extended to handle cubic cells (in addition to tetrahedra) as the thermal properties of biological tissues are assigned voxel-wise.}

\revOne{The system of optimality conditions associated to \eqref{eq:3D1Dopt_t}-\eqref{eq:3D1D_b_t} was solved via the preconditioned conjugate gradient (PCG) scheme reported in \cite[Algorithm 1]{berrone2022}, up to a relative residual of $10^{-6}$. 
Here, the scheme was adapted to efficiently handle the large linear systems characterizing the problems of interest. Each iteration of the PCG scheme, here called \textit{outer} iteration, requires the resolution of sub-systems to compute the descent direction. Each of these linear systems was solved by means of \textit{inner} iterations applying again a PCG scheme, until a relative residual of $10^{-7}$ is reached. To speed-up convergence, a preconditioner based on the incomplete Cholesky factorization is proposed for the \textit{inner} PCG loops, whereas the same preconditioner of \cite{berrone2022} is chosen for the \textit{outer} loop solver, here factorized to improve efficiency.
It is important to remark that the efficient and effective resolution of the inner sub-systems is key to avoid the assembly of the global coefficient matrix involved in the external PCG loop.}

As other 3D-1D coupling strategies, this approach does not require the 3D mesh to be conforming to the 1D inclusions, hence allowing for the use of a voxel-based mesh which is completely blind to the medical implant. Thanks to the introduction of the auxiliary variables $\phi$ and $\psi$, interface quantities are available at any stage of the simulation, without the need of post-processing.

\revOne{Finally, to measure how well the continuity condition at the interface between the implant and the tissue was enforced through minimization of the cost functional~\eqref{eq:functional}, the following relative discrepancy measure was used
\begin{equation*}
\Delta_\theta=\frac{||\vartheta_h-\hat \vartheta_{\hat h}||_\Gamma}{\|\vartheta_h\|_\Gamma}\,.
\end{equation*}}

\subsection{Experimental comparison}

The proposed numerical model was compared experimentally with temperature measurements acquired while exposing a \revOne{cranial mesh} embedded either in a gel phantom or in an expanded polystyrene phantom to the magnetic field generated by a system of MRI GCs.

\subsubsection{Experimental set-up}
\label{sec:experiment:setup}

\begin{figure}[!t]
  \centering
  \includegraphics[width=9cm]{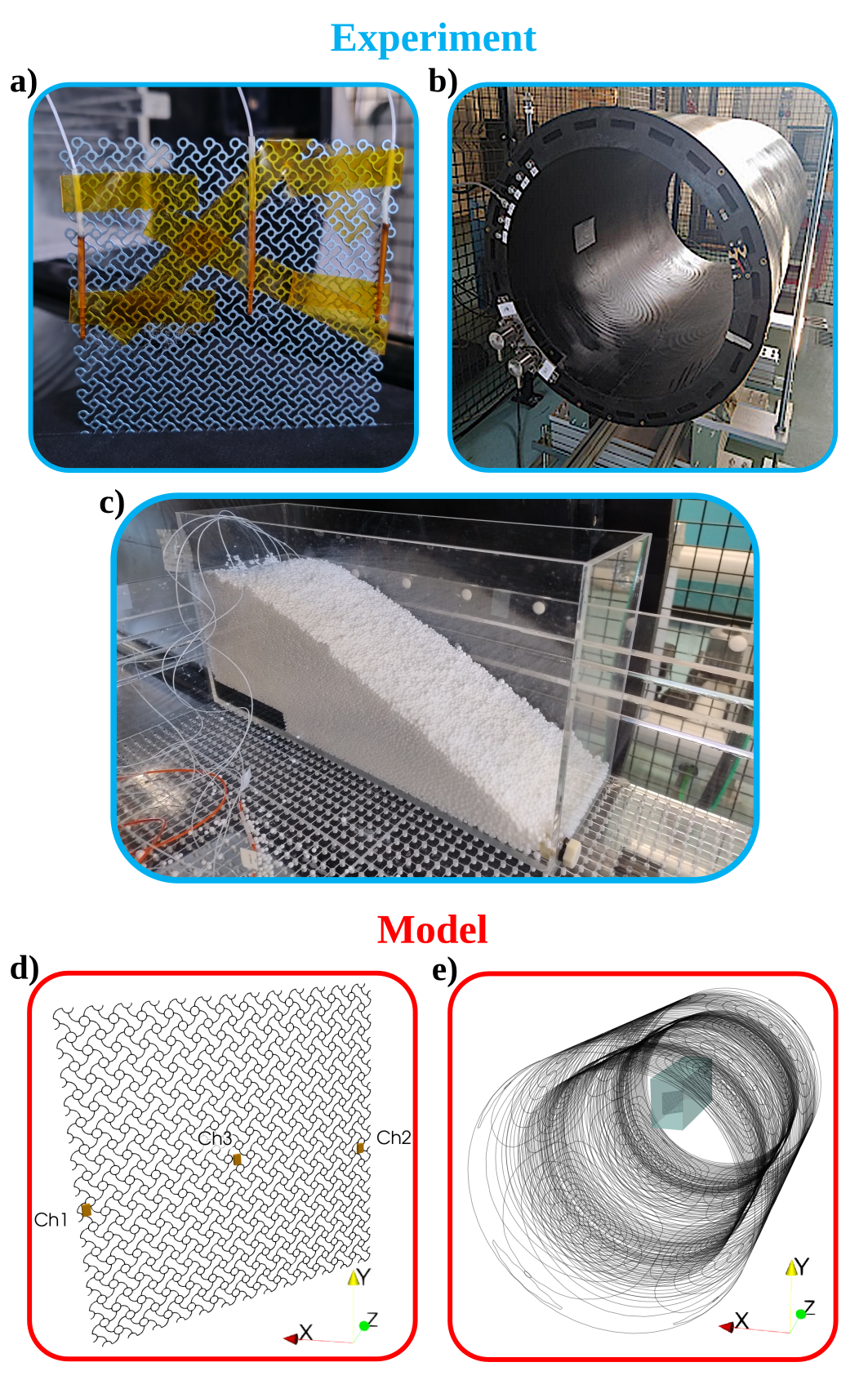}
  \caption{{\bf Experimental setup and virtual modelling.} {\bf a)} Semi-rigid \revOne{cranial mesh} in titanium equipped with the optical fibre temperature probes. {\bf b)} System of GCs for cylindrical MRI scanners. {\bf c)} Phantom filled with expanded polystyrene grains. {\bf d)} Computational model of the \revOne{cranial mesh} with depicted the boxes of the active parts of the temperature probes. {\bf e)} Computational domain including the phantom with the \revOne{cranial mesh} and the model of the GCs.}
  \label{fig:experiment-model}
\end{figure}

The measurements were conducted on a semi-rigid titanium (ASTM F67) \revOne{cranial mesh} manufactured by Medartis AG (Basel, Switzerland).
The implant was $\SI{93}{\milli\meter} \times \SI{93}{\milli\meter}$ and its thickness was \SI{0.6}{\milli\meter}. The thickness of the implant corresponds to the diameter of its one-dimensional structure, so that $r = \SI{0.3}{\milli\meter}$.
It was equipped with \revOne{three} optical fibre temperature probes\revOne{, two} positioned in the peripheral regions, where the largest temperature increase was expected to be induced\revOne{, and one in the central region, as shown in Fig.~\ref{fig:experiment-model}a}.
The probes were connected to an eight-channel AccuSens signal conditioner produced by Opsens (Québec, Canada) sampling at \SI{20}{\hertz}. The manufacturer declared an overall accuracy of $\pm \SI{0.30}{\degreeCelsius}$ and a resolution of \SI{0.01}{\degreeCelsius}.

In a first experiment, simulating the heat transfer from the \revOne{cranial mesh} towards soft tissues, the implant with the positioned probes were plunged into a cuboid container (base of $\SI{13}{\centi\meter} \times \SI{44}{\centi\meter}$, height of \SI{20}{\centi\meter}) filled with gel simulating average tissue parameters produced by Zurich MedTech AG (Zurich, Switzerland) in accordance with ISO/TS 10974~\cite{iso2018}.
In a second experiment, simulating an almost adiabatic condition, the same cuboid container was filled with expanded polystyrene grains whose diameter varies between \SI{1}{\milli\meter} and \SI{3}{\milli\meter} (Fig.~\ref{fig:experiment-model}c).
The base of the implant was fixed in a slot cut in a spongy support to keep it in the correct position within the container (see Fig.~\ref{fig:experiment-model}a). In particular, the implant was positioned at \SI{45}{\degree} with respect to the container walls to have it perpendicular to the generated magnetic field, which is the configuration expected to maximize the induced temperature increase.

The magnetic field was generated by the system of actively shielded GCs with whole-body access for cylindrical MRI scanners depicted in Fig.~\ref{fig:experiment-model}b. Precisely, it is the model Solaris-R manufactured by Nanjing Cichen Medical Technology Co., Ltd (Nanjing, China).
The system, featuring three coils to generate gradient magnetic fields along three orthogonal directions, has an internal diameter of about \SI{67}{\centi\meter} and a total length of about \SI{150}{\centi\meter}. The gradient directions determine the used reference system illustrated in Fig.~\ref{fig:experiment-model}e, with $\hat{z}$ directed longitudinally with respect to the coils, $\hat{y}$ directed vertically and $\hat{x}$ perpendicular to the other two directions.

The GCs were supplied by a NG500 1.3 gradient amplifier built by Prodrive Technologies (Eindhoven, The Netherlands) able to drive each coil independently and to provide a peak current of \SI{1000}{\ampere} and a maximum voltage of \SI{940}{\volt}.
To dissipate the loss power in the coil conductors due to the high current values, a water cooling system was connected to the coils. The cooling system made negligible the impact of the heating dissipated by the GCs' conductors on the temperature values measured on the implant, as verified through a \revOne{fourth} optical fibre temperature probe positioned at the boundary of the phantom, where no heating due to the currents induced on the \revOne{cranial mesh} was expected.
In the experiment, the coils generating gradients directed along $\hat{x}$ and $\hat{z}$ were driven with two in-phase sinusoidal currents at the frequency of \SI{2}{\kilo\hertz}.
According to the numerical model described in Section~\ref{sec:experiment:comparison}, the peak current intensities of \SI{150}{\ampere} were such that a peak magnetic field of \SI{3.5}{\milli\tesla} was generated at the implant barycentre, located at $x = \SI{14}{\centi\meter}$, $y = 0$ and $z = \SI{30}{\centi\meter}$ with respect to the coil isocentre (i.e., the central point where all the coils generate a null magnetic field).

The temperature detected by the optical fibre temperature probes were recorded every \SI{2}{\second} for \SI{900}{\second} of continuous exposure of the \revOne{cranial mesh} to the harmonic magnetic field generated by the GCs.
Temperature recording started \SI{30}{\second} before switching on the power amplifier to acquire the value at which the system and the probes were thermalised.
Since the actual measurand to be compared with the simulation results is the temperature increase, an offset equal to the average of the temperature values recorded in the preliminary \SI{30}{\second} were applied to the measurement results.

The measurements were repeated two times a day apart without moving neither the probe nor the phantom, to assess the repeatability of the experiment.

\begin{figure}[t]
\centering
  \includegraphics[width=\textwidth]{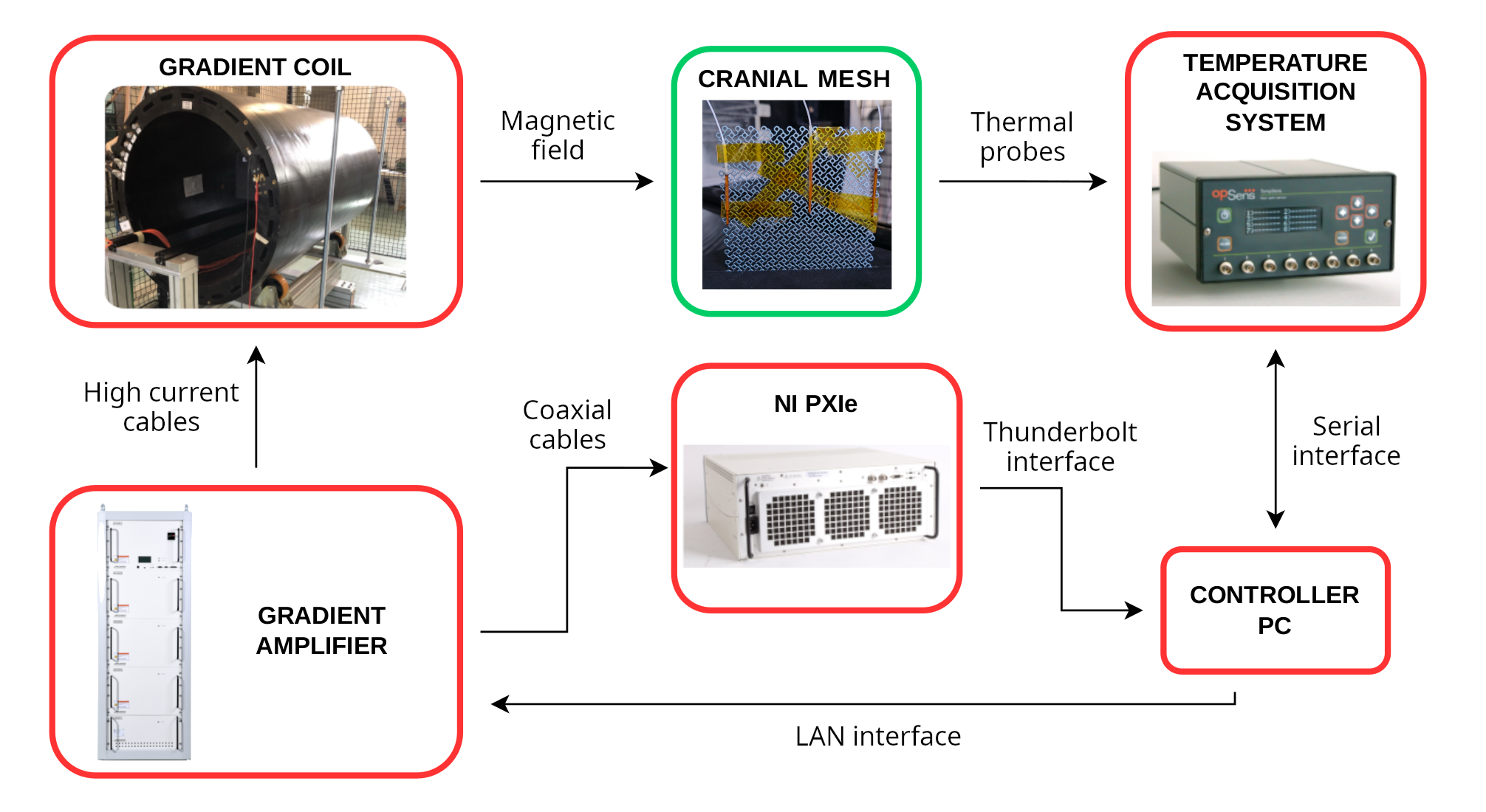}
  \caption{\revOne{{\bf Experiment pipeline.} The experiment was monitored by a controller PC that collected the acquired temperature data through a serial connection with the temperature acquisition system. The same controller PC controlled the gradient amplifier through a LAN connection and monitored the generated currents through a Thunderbolt connection and an acquisition board connected through coaxial cables to the gradient amplifier. The gradient amplifier supplied the GCs through high current cables and the \revOne{cranial mesh} was exposed to the generated magnetic field.}}
  \label{fig:experiment-pipeline}
\end{figure}

\revOne{Figure~\ref{fig:experiment-pipeline} depicts the measurement pipeline. The temperature probes attached to the implant were connected to the temperature acquisition system, which transmitted the acquired temperature data to the controller PC through a serial interface. Meanwhile, the currents flowing into the GCs were monitored through a 3-channel acquisition board connected to a National Instrument (Austin, Texas) PXIe. The current output monitor of the gradient amplifier were connected to the acquisition board with three coaxial cables and data from the acquisition board were transmitted to the PC through a Thunderbolt interface. The gradient waveform was set on the amplifier through a LAN interface controlled by the amplifier's proprietary software installed on the controller PC.}

\subsubsection{Numerical model}
\label{sec:experiment:comparison}

The one-dimensional model of the \revOne{cranial mesh} was created by repeating the elementary cell reported in Fig.~\ref{fig:cranial-plate}c until a square of \SI{93}{\milli\meter} sides was filled (see Fig.~\ref{fig:experiment-model}d). The entire implant geometry is described by \revOne{
about \num{8000}} segments.
The electrical currents induced in the implant were computed according to the method described in Section~\ref{sec:modelling:electromagnetic}, considering an electrical conductivity of \SI{1.82}{\mega\siemens\per\meter}.
The application of the Biot-Savart law to a model of the gradient coils (depicted in Fig.~\ref{fig:experiment-model}e) allowed the computation of the magnetic vector potential along the model branches, providing the electromotive forces to the electrical network problem.
The computed currents were processed according to~(\ref{eq:joule-law}) to evaluate the linear power density dissipated in the \revOne{cranial mesh}.

\begin{table}[b]
    \centering
    \caption{Thermal properties of the materials involved in the experimental comparison.}
    \label{tab:thermal-properties}
    \begin{tabular}{|c|c|c|c|}
        \hline
        Material & Thermal conductivity & Density & Specific heat capacity \\
         & (\si{\watt\per\meter\per\degreeCelsius}) & (\si{\kilo\gram\per\cubic\meter}) & (\si{\joule\per\kilo\gram\per\degreeCelsius}) \\
        \hline
        Gel phantom & 0.624 & 1006 & 4200 \\
        Expanded polystyrene & 0.035 & 20 & 1200 \\
        Titanium & 17 & 4510 & 523 \\
        \hline
    \end{tabular}
\end{table}

The evaluated dissipated power was used as the forcing term of the thermal problem, solved following both the simplified approach based on the thermal seeds and the complete 3D-1D coupled model.
For both the approaches, the computational domain was the cuboid phantom in which the \revOne{cranial mesh} was plunged into (see Fig.~\ref{fig:experiment-model}e). The domain was discretized with isotropic \SI{2}{\milli\meter} voxels and adiabatic conditions (i.e., $h_{\rm amb} = 0$) were imposed on its boundary.

The thermal properties of the gel phantom were provided by the vendor.
For the expanded polystyrene grains, the vendor declared a density of about \SI{20}{\kilo\gram\per\cubic\meter}, from which the thermal conductivity was deduced according to~\cite{gnipThermalConductivityExpanded2012}. Expanded polystyrene specific heat capacity was retrieved from~\cite{yousefiThermalConductivitySpecific2021}.
All the thermal property values used in the simulations are summarised in Table~\ref{tab:thermal-properties}.
Only in the 3D-1D coupled model, the thermal properties of titanium were used in the \revOne{cranial mesh}.

To compare the results of the numerical models with the measured values, the thermal probes were modeled by recreating their active parts as small boxes $\SI{3}{\milli\meter} \times \SI{2}{\milli\meter} \times \SI{1}{\milli\meter}$ positioned next to the \revOne{cranial mesh}. The exact positioning is illustrated in Fig.~\ref{fig:experiment-model}d, where the probes are depicted as orange boxes.

\subsection{Magnetic hyperthermia case study}
\label{sec:case-study}

As a test case, the proposed 3D-1D coupled model is compared to the approximated thermal seed model in the estimation of the heating of a \revOne{cranial mesh} in a patient undergoing a MH treatment for a neck superficial tumour.
To this end, the anatomical human model Glenn from the Virtual Population~\cite{gosselin2014} was equipped with the geometrical model of the semi-rigid \revOne{cranial mesh} described in Section~\ref{sec:experiment:setup}. The \revOne{cranial mesh} was properly deformed according to a transformation function $F$ from the plane to a curved surface in order to fit Glenn's skull (Fig.~\ref{fig:cranial-plate}a).
Moreover, a collar-type MH applicator operating at the frequency of \SI{100}{\kilo\hertz}, used for neck tumour treatment, is placed around the neck of Glenn~\cite{bottauscio2023}.

The MH applicator is driven to generate a maximum magnetic flux density within Glenn's body of about \SI{16}{\milli\tesla}, which satisfies the Atkinson--Brezovich limit with $H f \cong \SI{1.3e9}{\ampere\per\meter\per\second} < \SI{5e9}{\ampere\per\meter\per\second}$.
Hence, a patient without implants could undergo such a treatment.
However, the presence of the \revOne{cranial mesh} leads to an undesired temperature increase in proximity of the metallic implant that is here quantified through numerical simulations of a continuous exposure with \SI{900}{\second} duration.
In particular, the computations were performed on Glenn's head discretised with uniform \SI{1}{\milli\meter} voxels and assuming a thermal exchange with the environment described by $h_{\rm amb} = \SI{7}{\watt\per\square\meter\per\kelvin}$.
The implant was modelled with an electrical conductivity of \SI{1.82}{\mega\siemens\per\meter} and the thermal properties collected in Table~\ref{tab:thermal-properties}.
The thermal properties of the biological tissues were assigned in accordance to the IT'IS Foundation database~\cite{itis-database}.

\section{Results}
\label{sec:results}

\subsection{Experimental comparison}
\label{sec:experiment:measurement}

\begin{figure}[!t]
  \centering
  \includegraphics[width=12.85cm]{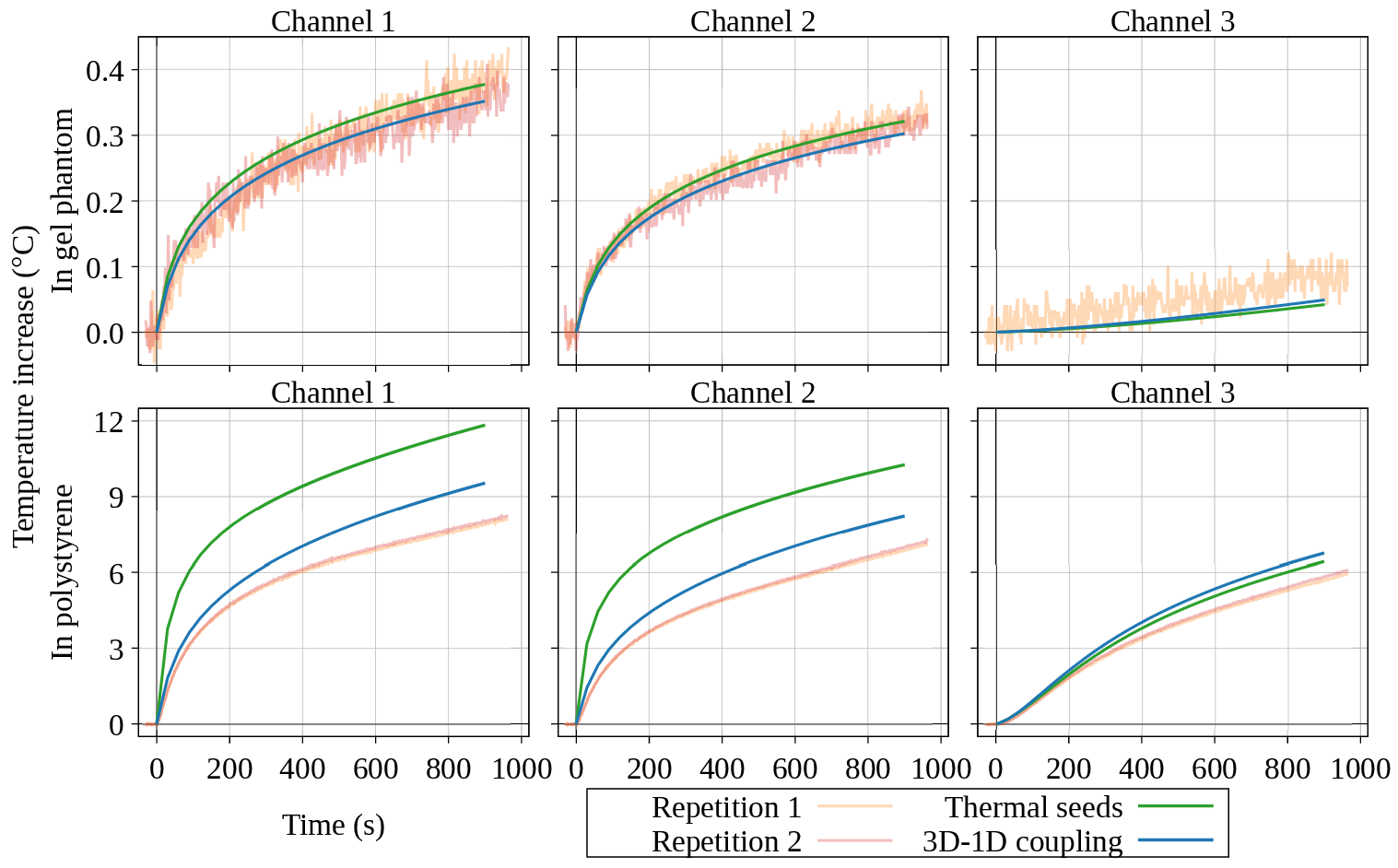}
  \caption{{\bf Comparison between measurements and simulations.} The noisy pale orange and pale red lines report the measurement results in the first and second repetition, respectively. The solid green and blue lines report the simulated results with the thermal seed model (purely 3D FEM) and the 3D-1D coupling, respectively. The results are reported for the \revOne{three} channels illustrated in Fig.~\ref{fig:experiment-model}d for both the gel phantom and the polystyrene. \revOne{During the second repetition of the experiment in gel, the third channel loosened its contact with the implant, leading to an untrustworthy measurement, whose result is not reported here.}}
  \label{fig:comparison-measurement}
\end{figure}

The results of the two repeated measurements are reported in Fig.~\ref{fig:comparison-measurement} and they agree with each other for both the gel phantom and the polystyrene.
\revOne{The comparison of repeated measurements is not possible for the probe attached near the centre of the implant (channel 3) in gel, because the probe loosened its contact with the implant during the second repetition, leading to an untrustworthy measurement, that is therefore not reported in Fig.~\ref{fig:comparison-measurement}.}
The small induced temperature increase in the case of the gel phantom makes the measurement noise a substantial source of measurement uncertainty for this experiment, whereas noise is not sensibly affecting the measurements in the phantom filled with expanded polystyrene grains.

The virtual measurements were performed by averaging the computed temperature increase within each box modelling the thermal probes and correspond to the values reported in Fig.~\ref{fig:comparison-measurement} along with the measurement results.

\begin{figure}[!t]
  \centering
  \includegraphics[width=9cm]{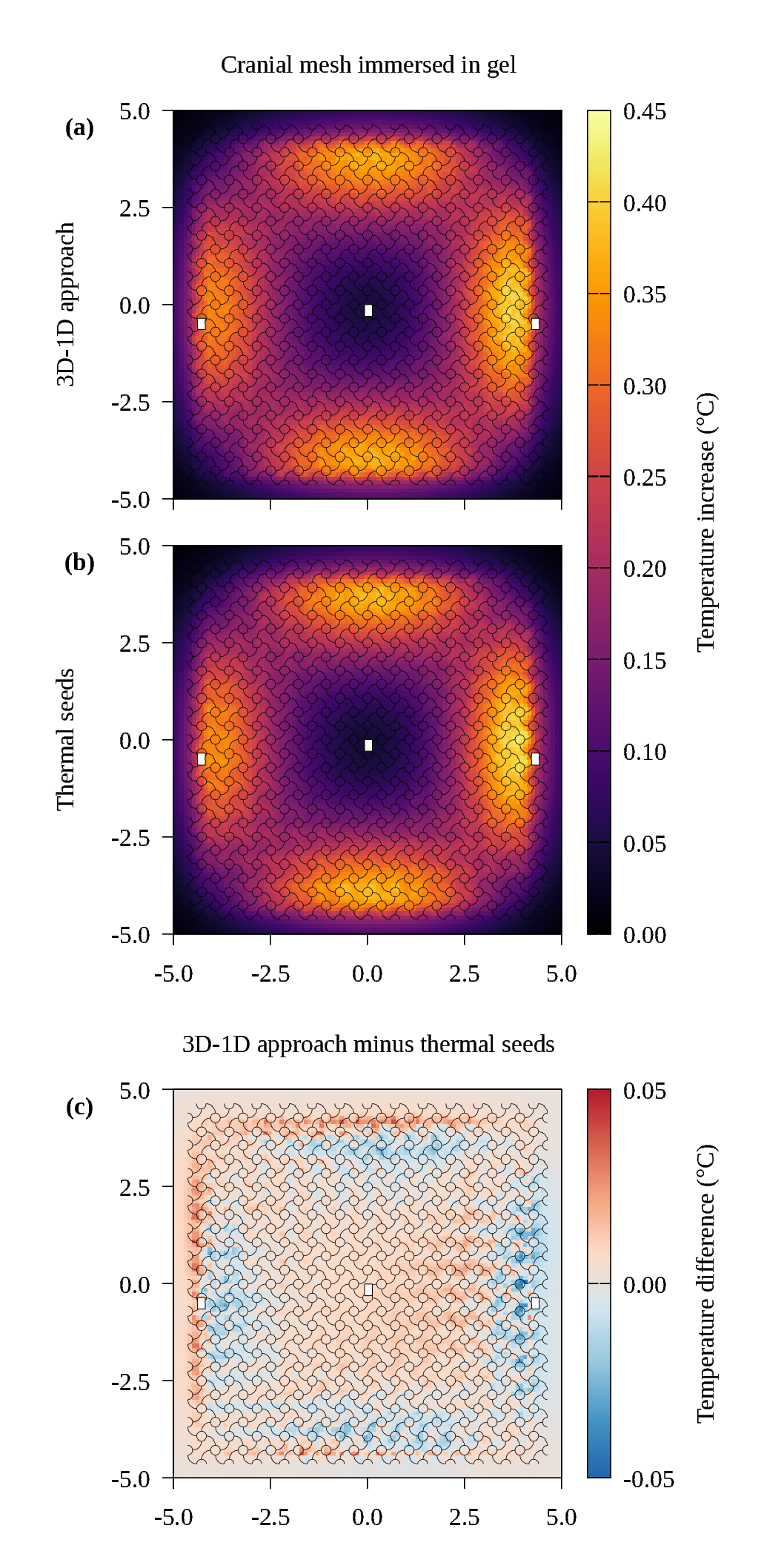}
  \caption{{\bf Temperature increase induced by MRI GCs on the \revOne{cranial mesh} immersed in gel.} {\bf a)} Result of the proposed approach based on 3D-1D coupling. {\bf b)} Result of the approximated thermal seed model (purely 3D FEM). {\bf c)} Difference between the results of the two models. In all the panels, horizontal and vertical axes are expressed in centimetres \revTwo{and the white rectangles represent the projections of the probes on the plane of the implant}.}
  \label{fig:temperature-in-gel}
\end{figure}

\begin{figure}[!t]
  \centering
  \includegraphics[width=9cm]{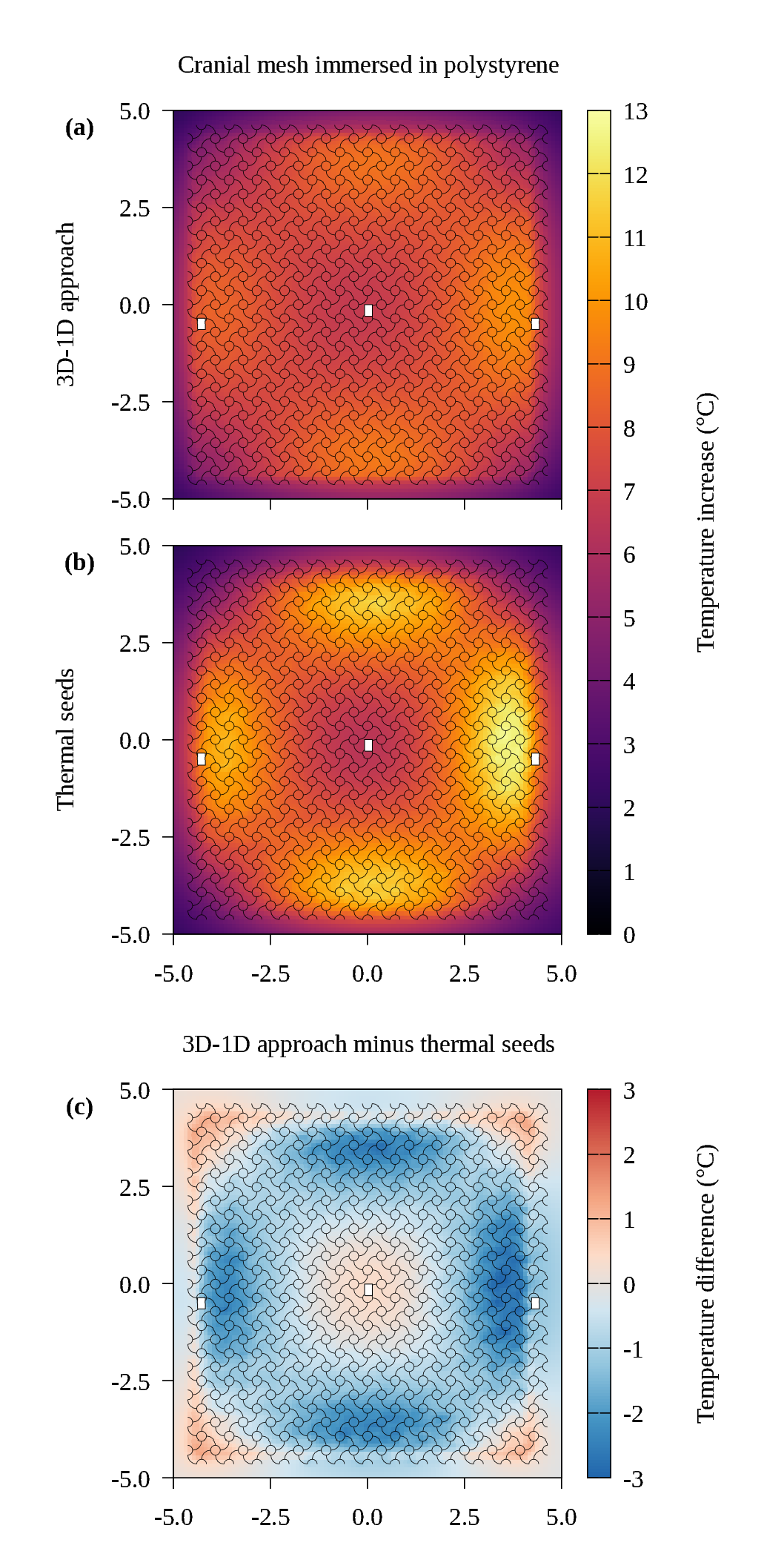}
  \caption{{\bf Temperature increase induced by MRI GCs on the \revOne{cranial mesh} immersed in polystyrene.} {\bf a)} Result of the proposed approach based on 3D-1D coupling. {\bf b)} Result of the approximated thermal seed model (purely 3D FEM). {\bf c)} Difference between the results of the two models. In all the panels, horizontal and vertical axes are expressed in centimetres \revTwo{and the white rectangles represent the projections of the probes on the plane of the implant}.}
  \label{fig:temperature-in-polystyrene}
\end{figure}

The distributions of the temperature increase estimated by the two approaches on the plane containing the \revOne{cranial mesh} are reported in Fig.~\ref{fig:temperature-in-gel}, for the phantom gel, and in Fig.~\ref{fig:temperature-in-polystyrene}, for the polystyrene grains. \revOne{In both cases, a relative continuity mismatch $\Delta_\vartheta$ of order \num{e-3} was measured for the 3D-1D approach.}

\subsection{Magnetic hyperthermia case study}
\label{sec:application}

\begin{figure}[!t]
  \centering
  \includegraphics[width=\textwidth]{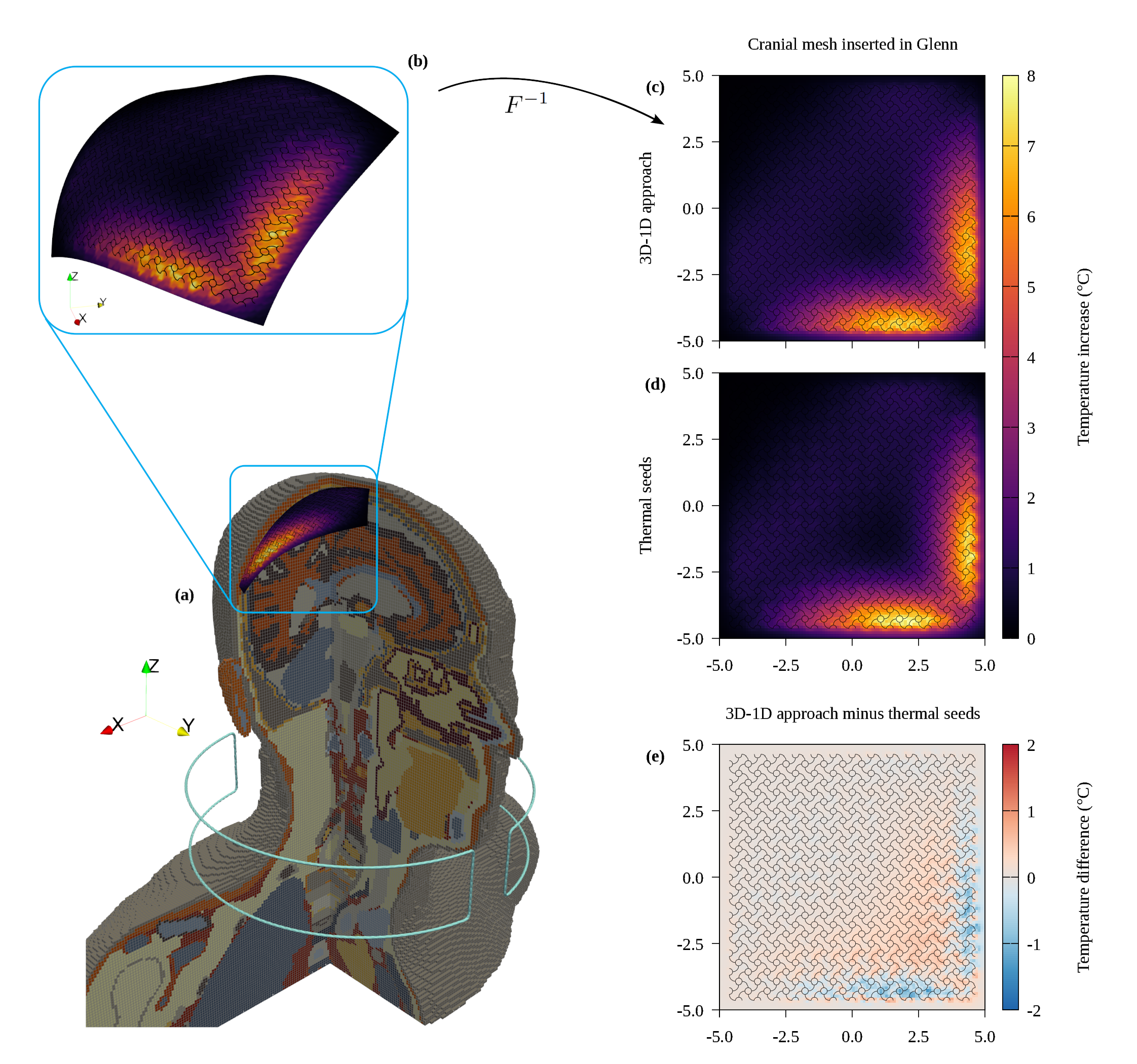}
  \caption{{\bf Temperature increase induced by a MH applicator on the \revOne{cranial mesh} implanted in Glenn.} {\bf a)} Anatomical human model Glenn with the implanted \revOne{cranial mesh} and the collar-type MH applicator for neck tumour treatment. {\bf b)} A surface containing the implanted \revOne{cranial mesh} is extracted and deformed into a plane. On this plane are reported: {\bf c)} the result of the proposed approach based on 3D-1D coupling; {\bf d)} the result of the approximated thermal seed model (purely 3D FEM); and {\bf e)} the discrepancy between the results of the two models. In all the panels, horizontal and vertical axes are expressed in centimetres.}
  \label{fig:temperature-in-hyperthermia}
\end{figure}

In order to compare the temperature increase distributions provided by the two models in proximity of the \revOne{cranial mesh}, the temperature increase values computed on the curved surface containing the \revOne{cranial mesh} were extracted. The surface was then deformed back to a plane according to the inverse transformation function $F^{-1}$ (see Fig.~\ref{fig:temperature-in-hyperthermia}a and \ref{fig:temperature-in-hyperthermia}b). The temperature increase distributions estimated by the 3D-1D approach and by the thermal seed model on the resulting plane are reported in Fig.~\ref{fig:temperature-in-hyperthermia}c and Fig.~\ref{fig:temperature-in-hyperthermia}d, respectively.
The difference between the two temperature distributions is reported in Fig.~\ref{fig:temperature-in-hyperthermia}e. \revOne{Also in this case, a relative continuity mismatch $\Delta_\vartheta$ of order \num{e-3} was measured for the 3D-1D approach.}

\section{Discussion}
\label{sec:discussion}

\subsection{Experimental comparison}

According to Fig.~\ref{fig:comparison-measurement}, in the case of the gel phantom, both the approximated model based on thermal seeds and the complete 3D-1D coupled model agree with the noisy measurement results. The numerical model and the computational procedure handling the 3D-1D coupling are, therefore, experimentally validated, as well as the thermal seed approximation.
Interestingly, the 3D-1D coupled model evaluates a temperature increase detected by the thermal probes \revOne{in the peripheral region (channels 1 and 2)} that, after \SI{900}{\second} of exposure, is about the \SI{7}{\percent} lower than the temperature increase estimated by the thermal seed model.
\revOne{At the same time, the temperature increase detected by the thermal probe near the centre of the implant (channel 3) according to the 3D-1D coupled model is slightly larger than that estimated by the termal seed model.}
This happens because of the proper modelling of the thermal conductivity within the \revOne{cranial mesh}, along its thin but highly conductive branches.

Indeed, the dissipated power is mostly located \revTwo{near the edges} at the boundary of the implant. Therefore, only part of the generated heat is transferred to the surrounding gel; the remaining part of the heat stays within the implant and moves towards the inner \revTwo{and corner regions}, where no power is directly dissipated by the induced electrical currents. This secondary physical effect, that tends to uniform the temperature in the implant \revTwo{by reducing its value in the hot-spots and increasing it elsewhere}, cannot be described by the approach based on thermal seeds, whereas it is modeled by the 3D-1D approach, as highlighted in the map of the difference between the two estimated distributions in Fig.~\ref{fig:temperature-in-gel}c.

The second experiment, with the \revOne{cranial mesh} plunged into the expanded polystyrene grains, led to a larger temperature increase and a clearer separation between the results obtained from the two models \revOne{in channels 1 and 2}. From the results reported in Fig.~\ref{fig:comparison-measurement}, the greater accuracy of the 3D-1D coupling approach with respect to the thermal seed model in describing the actual physical phenomenon is clear.
After \SI{900}{\second} of exposure, the temperature increase computed with the 3D-1D coupled model is about \SI{25}{\percent} lower than the temperature increase estimated with the thermal seed model \revOne{in both the peripheral probes (channels 1 and 2)}.
Moreover, the trend of the temperature increase computed with the 3D-1D coupled model is in close agreement with the measurement results, especially during the first \SI{100}{\second} of the experiment.
Many reasons can motivate the separation between the two trends after this time period, like the imperfect modelling of the probe active part, as well as the thermal exchange with the environment through the boundary of the computational domain, and the not perfectly known thermal properties of the expanded polystyrene grains.
Nonetheless, the reached agreement between experimental measurements and simulations of the proposed 3D-1D coupled model is satisfactory and proves its larger accuracy (i.e., its capability to better reproduce the experimental behaviour) with respect to the thermal seed model.

\revTwo{The discrepancy between measurements and simulations is also found in the results collected by the probe located in the center of the implant (channel 3), where both the numerical models overestimated the measured values, although the agreement is larger than for the other probes.
Although for this probe the two simulations lead to very similar results, making impossible to discriminate between them in terms of accuracy with respect to the measured trend, it is worth noting that the 3D-1D coupled model estimated a temperature increase about \SI{5}{\percent} larger than the thermal seed model.
This happens because of the heat transfer through the highly conductive metal of the implant, that is properly modelled by the 3D-1D coupling.}

\revTwo{This fact becomes clearer by looking at the difference between the two temperature distributions estimated with the proposed method and the thermal seed approximation in Fig.~\ref{fig:temperature-in-polystyrene}c.
The difference is significantly larger than in the case of the gel phantom because the temperature increase computed with the 3D-1D approach is quite homogeneous within the entire implant (Fig.~\ref{fig:temperature-in-polystyrene}a), whereas the result of the thermal seed model is strongly heterogeneous, directly reflecting the distribution of the Joule losses due to the induced electrical currents (Fig.~\ref{fig:temperature-in-polystyrene}b).
Moreover, the map of the difference reported in Fig.~\ref{fig:temperature-in-polystyrene}c shows that near the vertices of the implant the temperature increase estimated using the 3D-1D approach is significantly larger than the values estimated with the thermal seeds. Precisely, near the vertices the thermal seed approximation underestimates the temperature increase of about \SI{1.3}{\celsius} with respect to the 3D-1D coupled model, with a relative discrepancy of almost \SI{35}{\percent}.}

\subsection{Magnetic hyperthermia case study}

From the results of both the 3D-1D coupling model (Fig.~\ref{fig:temperature-in-hyperthermia}c) and the thermal seed approximation (Fig.~\ref{fig:temperature-in-hyperthermia}d), it appears clear that the electrical currents induced by the MH applicator on the \revOne{cranial mesh} are strongly non uniform, leading to a heat deposition strongly focused on \revTwo{the} corner of the implant \revTwo{nearest to the right ear of Glenn}.

Because of the lack of thermal diffusion through the metal, the maximum temperature increase value is overestimated by the thermal seed model, according to which a peak temperature increase of about \SI{8.2}{\celsius} is reached next to the implant.
A less conservative and more accurate maximum temperature increase value of about \SI{7.4}{\celsius} is estimated by the 3D-1D coupled model, that is about \SI{10}{\percent} lower than the result of the thermal seed model.
Despite the lower peak value, it can be noticed that the heated region computed by the 3D-1D coupled model is more extended than the one computed by the thermal seed model, as it can be appreciated also from the difference map reported in Fig.~\ref{fig:temperature-in-hyperthermia}e. This is a consequence of the heat diffusion through the highly conductive metal of the \revOne{cranial mesh}, which is properly taken into account in the 3D-1D coupled model, \revTwo{although the effect is not so remarkable as in the simulation of the implant immersed in the polystyrene grains}.

\section{Conclusions}
\label{sec:conclusions}

The aim of this paper was to quantify reliably the heating induced in a patient implanted with a thin, one-dimensionally structured passive implant when exposed to time-varying magnetic fields.
This goal has been reached by adopting an innovative mathematical and numerical modelling strategy in which a three-dimensional problem describing the heat transfer in the biological tissues is coupled with a one-dimensional problem describing the heat transfer in the metallic implant.
In this paper, the developed method has been tested by assessing the induced temperature increase in a cranial mesh exposed to the magnetic fields generated by an MRI GC system and a MH applicator, \revTwo{but the same procedure could be directly applied to other medical devices, like metallic stents~\cite{bottauscio2022} or orbital implants~\cite{sullivan1994},} \revOne{and other magnetic field sources, like wireless power transfer systems~\cite{arduino2020}.}

The proposed model has been compared with experimental measurements showing that, by taking into account the heat transfer through the metallic implant, the proposed model reaches more accurate and less conservative estimations \revTwo{of the maximum temperature increase} than the previously adopted approximated thermal seed model.
The overestimation of the thermal seed model with respect to the results of the proposed 3D-1D coupled model varied from \SI{7}{\percent}, in the case of the \revOne{cranial mesh} embedded in gel phantom, up to \SI{25}{\percent}, in the case of the phantom filled with expanded polystyrene grains.
In the MH case study, where an anatomical human model was simulated, the thermal seed model overestimated the maximum temperature increase value by about \SI{10}{\percent}.

\revTwo{Since it overestimates the maximum induced temperature increase, the thermal seed approximation could be seen as a simple and conservative way for assessing the safety of a scenario under investigation.
In reality, the adoption of the thermal seed approximation in place of the 3D-1D coupled model comes with a number of drawbacks.
On the one hand, employing an excessively conservative method to evaluate the expected maximum induced temperature increase and assess the safety of a medical procedure could limit inappropriately the access to healthcare to implanted people.
On the other hand, estimating a wrong temperature distribution, with significant local underestimations, like the \SI{35}{\percent} one observed in the simulation of the cranial mesh immersed in the polystyrene grains, could lead to underestimated risks anytime the safety assessment is based on the entire temperature distribution, instead of its maximum value only.}

\revTwo{The latter case is often of practical relevance, making the proposed 3D-1D coupled model important for the proper modelling of the heat transfer within the metallic implant. For example, the actual distribution of the induced temperature increase is relevant in all the cases in which the implant is in contact with different tissue pockets, each with a different sensitivity to heat, leading to a thermal risk profile that is not homogeneous across the implant.
In addition, the actual distribution is relevant any time the heating due to the eddy currents induced in the implant are not the only source of heating, like in MH where the temperature increase due to the eddy currents induced in the implant should be combined with the one due to the eddy currents induced in the tissues and to the losses induced in the MNPs. Similarly, in MRI the combined effect of the magnetic field generated by the GCs and the RF field could make relevant the entire temperature distribution~\cite{arduino2021} for a comprehensive safety assessment.}

\section*{Acknowledgment}
The results presented here were partially developed in the framework of the 21NRM05 STASIS project. This project has received funding from the European Partnership on Metrology, co-financed from the European Union's Horizon Europe Research and Innovation Programme and by the Participating States.
The specimen of the \revOne{cranial mesh} was kindly provided by the manufacturer Medartis (\url{www.medartis.com}).

Author Denise Grappein acknowledges to be holder of a Postdoctoral fellowship financed by INdAM (Istituto Nazionale di Alta Matematica) and hosted by the Research Unit of Politecnico di Torino. Authors Denise Grappein, Stefano Scialò and Fabio Vicini are members of the GNCS-INdAM group.

\section*{Data statement}
Data and codes used to generate the plots of the paper are available at doi:~\href{https://doi.org/10.5281/zenodo.16788714}{10.5281/zenodo.16788714}.

\bibliography{bibliography}

\end{document}